# Using an incoherent target-return to adaptively focus through atmospheric turbulence

W. Nelson[1,*], J. P. Palastro[2], C. Wu[1], and C.C. Davis[1]

[1]Department of Electrical and Computer Engineering, University of Maryland
College Park, Maryland 20742
[2]Naval Research Laboratory, Plasma Physics Division
Washington, DC 20375-5346

*Corresponding author: wnelson2@umd.edu

A laser beam propagating to a remote target through atmospheric turbulence acquires intensity fluctuations. If the target is cooperative and provides a coherent return beam, the phase measured near the beam transmitter and adaptive optics can, in principle, correct these fluctuations. Generally, however, the target is uncooperative. In this case, we show that an incoherent return from the target can be used instead. Using the principle of reciprocity, we derive a novel relation between the field at the target and the reflected field at a detector. We simulate an adaptive optics system that utilizes this relation to focus a beam through atmospheric turbulence onto the incoherent surface.

**Keywords**: atmospheric turbulence, reciprocity, adaptive optics, scattering

A laser beam propagating through atmospheric turbulence acquires phase aberrations from turbulent fluctuations in the refractive index. These aberrations can quickly develop into intensity fluctuations degrading the beam profile incident on a remote target [1-5]. In principle, adaptive optics can correct for the phase aberrations if the turbulent channel is reciprocal, a property of optical configurations in which the point spread function is invariant with respect to interchange of the source and receiver [5-7]. One adaptive optics configuration utilizes a cooperative target that provides a spatially coherent emitter return from a point on or near the target. This is essentially a beacon or guide star [5,8]. A wavefront sensor, placed near the beam transmitter, measures the phase of light received from this emitter. If the conjugate phase is applied to the transmitted beam, it reproduces the emitter's irradiance profile at the target. In the special case the target emitter is point-like, the measured phase provides the turbulent channel's point spread function, and the target can be illuminated with a desired irradiance profile in the vicinity of the emitter.

Often there is no coherent emitter at or near the target, the state of the turbulent channel remains unknown, and one must rely on a reflection from the target. For normal-incidence mirror reflections, the light received at the wavefront sensor has passed through the turbulent channel twice, and the phase aberration acquired in each pass cannot be explicitly determined. Scattering from a rough surface add a further complication. These surfaces scramble the beam phase such that the reflected beam loses its spatial coherence. In spite of these complications, we show here that an adaptive optics system can use the incoherent return to correct phase aberrations and focus the beam onto the target. In particular, we use the principle of reciprocity to derive a novel relation between the field incident upon a reflector and the field detected by a point receiver. The derivation considers a double pass through the turbulent channel: propagation from the source to the target and back to the detector, making it especially suitable for uncooperative targets. Simulations of an adaptive optics system using the relation demonstrate enhancements in both the on-target intensity and power in bucket.

We begin by using reciprocity to derive a relation between the field incident upon a reflector and the field detected by a point receiver. While the relation is general to any reciprocal optical configuration, we discuss it here in the context of focusing a laser beam through atmospheric turbulence onto a target that provides a spatially incoherent return. Figure 1 displays the reciprocity preserving geometry considered throughout. The beam starts at the transmitter plane propagates through the turbulent channel, and undergoes scattering at the target plane. The scattered beam is collected in the receiver plane, where it is focused to the detector. The transmitter and receiver planes coincide, creating a monostatic channel. An aperture and phase modulator are located in the transmitter/receiver plane. A thin lens with focal length $f$ is centered a distance $f$ from the detector and transmitter/receiver planes.

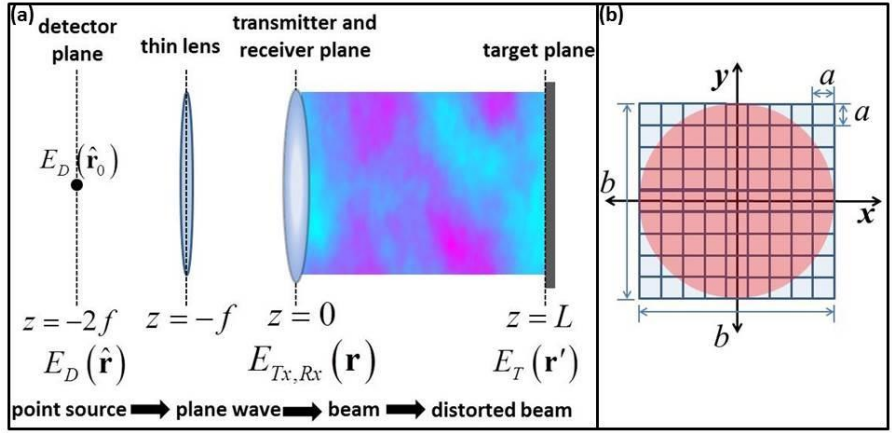

Figure 1. Schematic of the propagation geometry (a) and phase modulator (b). The phase modulator is located in the transmitter and receiver plane.

We express the transverse electric field, $E_\perp$, as a carrier wave modulating a slowly varying envelope, $E$: $E_\perp(\mathbf{x},t) = \frac{1}{2} E(\mathbf{x}) \exp[i(kz - \omega t)] + \text{c.c.}$ where $k = \omega/c$. Since we are only interested in the electric field envelope at specific planes, we adopt the shorthand $E_D(\hat{\mathbf{r}})$, $E_O(\mathbf{r})$, $E_R(\mathbf{r})$, and $E_T(\mathbf{r}')$ for the complex electric field envelopes at the detector, transmitter, receiver, and target respectively, where $\hat{\mathbf{r}}$, $\mathbf{r}$, and $\mathbf{r}'$ are the associated transverse coordinates. The spatial coordinates and fields are normalized by $(2\pi f / k)^{1/2}$ and $(2I_O / \varepsilon_0 c)^{1/2}$ respectively, where $I_O$ is the peak intensity of the outgoing beam. The function $\Theta(\mathbf{r})$ describes the transfer function of the aperture and phase modulator in the transmitter/receiver plane, which can be any complex, square integrable function.

The initial envelope of the transmitted beam can be expressed as the field of a point source located at $\hat{\mathbf{r}}_0$ propagated from the detector plane to the transmitter plane and filtered by the function $\Theta(\mathbf{r})$:

$$E_O(\mathbf{r}) = -i\Theta(\mathbf{r}) \exp(-2\pi i \hat{\mathbf{r}}_0 \cdot \mathbf{r}). \qquad (1)$$

We denote the point spread function, or Green's function for propagation from the transmitter to the target as $G_{TO}(\mathbf{r}';\mathbf{r})$. In terms of the Green's function, the electric field envelope at the target is then given by

$$E_T(\mathbf{r}') = \int G_{TO}(\mathbf{r}';\mathbf{r}) E_O(\mathbf{r}) d\mathbf{r}, \qquad (2)$$

which upon making use of Eq. (1) provides

$$E_T(\mathbf{r}') = \int G_{TO}(\mathbf{r}';\mathbf{r})\Theta(\mathbf{r})\exp(-2\pi i\hat{\mathbf{r}}_0 \cdot \mathbf{r})d\mathbf{r} \ . \tag{3}$$

The target reflects the field, and in general both apertures the beam and imparts a phase shift, such that the reflected field has the initial profile $E_T(\mathbf{r}')F(\mathbf{r}')\exp[i\phi(\mathbf{r}')]$, where $F$ is a real square-integrable function and $\phi$ is real. Denoting $G_{RT}(\mathbf{r};\mathbf{r}')$ as the Green's function for propagation from the target to the receiver, we have

$$E_R(\mathbf{r}) = \int G_{RT}(\mathbf{r};\mathbf{r}')E_T(\mathbf{r}')F(\mathbf{r}')e^{i\phi(\mathbf{r}')}d\mathbf{r}' . \tag{4}$$

We stress that $\phi(\mathbf{r}')$ can represent any phase shift imparted upon reflection.

Back at the receiver plane, the field passes through the aperture and phase modulator and propagates from the front focal plane of the lens to the back focal plane, coinciding with the detector plane. The resulting envelope at the detector is then

$$E_D(\hat{\mathbf{r}}) = -i\int \Theta(\mathbf{r})\exp(-2\pi i\hat{\mathbf{r}} \cdot \mathbf{r})E_R(\mathbf{r})d\mathbf{r} , \tag{5}$$

Using Eq. (4) we can rewrite this as

$$E_D(\hat{\mathbf{r}}) = -i\int E_T(\mathbf{r}')F(\mathbf{r}')e^{i\phi(\mathbf{r}')}\left[\int G_{RT}(\mathbf{r};\mathbf{r}')\Theta(\mathbf{r})\exp(-i2\pi\hat{\mathbf{r}} \cdot \mathbf{r})d\mathbf{r}\right]d\mathbf{r}' . \tag{6}$$

If the channel between the transceiver and target is reciprocal, the Green's function for propagation from the transmitter plane to the target plane is equal to the Green's function for propagation from the target plane to the receiver plane [9,10]: $G_{TO}(\mathbf{r}';\mathbf{r}) = G_{RT}(\mathbf{r};\mathbf{r}')$. If we observe the field at the location $\hat{\mathbf{r}}_0$ we can write Eq. (6) in terms of the envelope at the target plane described by Eq. (3), which gives the final result

$$E_D(\hat{\mathbf{r}}_0) = \int E_T^2(\mathbf{r}')F(\mathbf{r}')e^{i\phi(\mathbf{r}')}d\mathbf{r}' . \tag{7}$$

Recall that the location $\hat{\mathbf{r}}_0$ is the location of the point source corresponding to the transmitted beam and can be determined by the tilt of the transmitted beam. Experimentally, this location can be found through the enhanced backscatter phenomenon [11].

Equation (7) relates a single quantity measured by the detector, $E_D(\hat{\mathbf{r}}_0)$, to the field at the target plane $E_T(\mathbf{r}')$, with information about the optical configuration contained only implicitly in $E_T(\mathbf{r}')$. The remaining factor, $F(\mathbf{r}')\exp[i\phi(\mathbf{r}')]$, captures the interaction with the target, and can take any form. Reflection from a mirror at normal incidence, for instance, is described by $F(\mathbf{r}')=1$ and $\phi(\mathbf{r}')=0$, while an idealized glint could be described by $F(\mathbf{r}') = A_g\delta(\mathbf{r}'-\mathbf{r}'_g)$, where $\mathbf{r}'_g$ is the origin of the glint on the target and $A_g$ its effective area. Of primary interest here is reflection from a rough surface, which can be described by a pointwise random phase $\phi(\mathbf{r}') \sim \text{unif}[0, 2\pi]$ with $F(\mathbf{r}')=1$. At first glance, it might appear that application of such a phase would eliminate implicit information of the optical configuration contained in Eq. (7): with $E_T = 1$, $E_D(\hat{\mathbf{r}}_0)$ averages to zero. However, an interpretation of Eq. (7) in terms of a random walk in the complex plane reveals its significance.

We conceptualize the continuous target surface as a grid of discrete scatterers. Each scatterer is separated by a distance $l$, occupies an area $l^2$, and has a uniformly distributed random phase. This is equivalent to approximating Eq. (7) as a Riemann sum:

$$E_D(\hat{\mathbf{r}}_0) = \sum_{n=1}^{N} \alpha_n e^{i\varphi_n} , \qquad (8)$$

where $\alpha_n \propto |E_T(\mathbf{r}'_n)|^2 \, l^2$, $\varphi_n = 2\arg[E_T(\mathbf{r}'_n)] + \phi(\mathbf{r}'_n)$, $\mathbf{r}'_n$ is the location of the n$^{th}$ scatterer, $N$ is the number of scatterers illuminated by the incident radiation, and we have mapped the 2D coordinate system to a 1D array for convenience. Equation (8) is a sum over random phasors, and can thus be interpreted as a random walk in the complex plane [12]. Typically the transverse coherence length of optical phase fluctuations [13] far exceeds the correlation length of a rough surface [14]. We thus ignore the phase contributed by the incident field in the following discussion such that $\varphi_n \sim \text{unif}[0, 2\pi]$. Simulations presented below provide further justification for this.

The magnitude of each step in the random walk is proportional to the intensity at a single location in the target plane, $\alpha_n$, while the direction of each step is determined by $\varphi_n$. The walk's result is the field measured in the detector plane. For simplicity, we consider the uniform illumination of an incoherent reflector, corresponding to an isotropic, uniform random walk. This fixes all $\alpha_n$ to a single value, $\alpha_n = \alpha$, such that the expected intensity in the detector plane is given by $I_E = <|E_D(\hat{\mathbf{r}}_0)|^2> = \alpha^2 N$, where the brackets denote the expected value [15].

Because $\varphi_n$ is determined by the target, it cannot be modified, and the random walk will remain isotropic. However, by adjusting the outgoing beam's phase, the intensity distribution on the target, i.e. the length of each step in the walk, can be modified. In particular, if the beam power, $P \propto \alpha N$, is fixed, then $\alpha N = \text{const.}$, and $I_E \propto \alpha P$. This proportionality demonstrates that, on average, an intensity increase at the detector point $\hat{\mathbf{r}}_0$ translates to an increase in the target plane intensity.

We continue by applying this analysis to focusing a laser beam through atmospheric turbulence onto a spatially incoherent reflector. Specifically, we simulate the optical configuration depicted in Fig. 1. A $\lambda = 2\pi/k = 1$ $\mu$m, collimated laser beam is initialized in the transmitter plane with a transverse Gaussian profile of spotsize $w_0$. Between the transmitter and the target, the beam evolves according to the steady-state paraxial wave equation with turbulence-induced refractive index fluctuations included through phase screens [16,17]. In the phase screen approximation, the beam acquires the accumulated phase distortions due to turbulence at discrete axial points along the propagation path, and is propagated in vacuum between these points. We use an analytic approximation of the Hill spectrum to characterize the distribution of turbulent scale sizes with an inner scale of 1 mm and outer scale of 1 m [4].

To simulate scattering from the rough surface at the target plane, we first apply a random phase uniformly distributed from zero to $2\pi$ to the electric field amplitude at each grid point [12,18]. We then apply a circular filter in the Fourier domain that eliminates transverse wave numbers near the domain boundary. This removes transverse wavenumbers that will leave the spatial domain upon propagation and ensures that all directions are limited by the same transverse wavenumber. Propagation between the receiver and detector plane is modeled using a Fourier transform and appropriate scaling of the domain lengths based on the focal length of the lens.

As a demonstration of Eq. (7) and to verify the numerical model, we simulated $w_0 = 5$ cm beam propagation through 1000 realizations of the optical configuration shown in Fig. (1). The turbulent channel was 2.5 km long, and composed of 10 phase screens each with a refractive index structure function, $C_n^2$, chosen randomly from a uniform distribution on the interval $[10^{-16}, 10^{-13}]$ m$^{-2/3}$. Choosing $C_n^2$ randomly for each realization provides verification over a wide range of turbulence-induced beam distortions in both the detector and target planes. The aperture function was set to a Gaussian, $\Theta(\mathbf{r}) = \exp(-r^2/w_0^2)$. Figure 2 displays scatter plots of the resulting left and right hand sides of Eq. (7) using the previously described normalization. Both the phase and amplitude predicted by the simulation are in excellent agreement with Eq. (7). We note that the simulation includes the turbulence phase contribution neglected in the random walk discussion above.

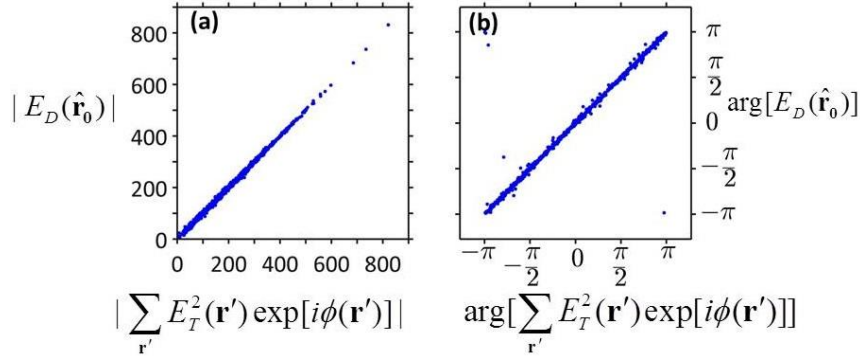

Figure 2. Scatter plot of the magnitude (a) and phase (b) depicting the numerical equivalent of the relation described by Eq. (7).

The random walk interpretation of Eq. (7) suggested a positive correlation between the focal and target intensities. Thus a feedback loop where the focal intensity informs the phase imparted to the transmitted beam, should, on average, result in an increased intensity on target. In the remainder, we describe simulations of an adaptive optics implementation that does just that. The adaptive optics system consists of a phase modulator (Fig. 1) in the transmitter/receiver plane controlled by the stochastic parallel gradient descent algorithm (SPGD), which we return to below. The phase modulator is comprised of square elements length $a$ on a side in a $15 \times 15$ array. Each element can make piston phase changes. The elements are indexed by $l$ and $m$ corresponding to the $x$ and $y$ dimensions respectively, and are defined on the interval $[-7, 7]$. For example, the indices $(l = -7, m = -7)$, $(l = 0, m = 0)$, and $(l = 7, m = 7)$ correspond to the bottom left, center, and top right elements respectively.

We use the conventional stochastic parallel gradient descent (SPGD) algorithm and notation [19]. The SPGD algorithm is designed to iteratively increase a metric by applying random perturbations to all control parameters and calculating the gradient of the metric. Motivated by the preceding discussion, our metric is $J = |E_D(\hat{\mathbf{r}}_0)|$. The phases applied by each element of the phase modulator serve as the control parameters. In particular, the phase applied by element $l, m$ at the end of iteration $n$ is defined as $u_{l,m}^{(n+1)}$ calculated in three steps:

$$u_{l,m}^{(+)} = u_{l,m}^{(n)} + \varepsilon S_{l,m}, \tag{9a}$$

$$u_{l,m}^{(-)} = u_{l,m}^{(n)} - \varepsilon S_{l,m}, \tag{9b}$$

$$u_{l,m}^{(n+1)} = u_{l,m}^{(n)} + \gamma S_{l,m}\left(J_+^{(n)} - J_-^{(n)}\right), \tag{9c}$$

where $S_{l,m}$ is randomly chosen as +1 or -1 for each element, $\varepsilon$ is the magnitude of the perturbation, $J_{+/-}^{(n)}$ corresponds to the metric value resulting from $u_{l,m}^{(+/-)}$, and $\gamma$ is the gain coefficient.

For the simulation, we chose the adaptive optics parameters $a = 2$ cm, $\varepsilon = \pi/10$, and $\gamma = 6\times10^{-4}$. The turbulent channel was 2.5 km in length with a $C_n^2 = 10^{-15}$ m$^{-2/3}$. This is considered mild optical turbulence; we elaborate on the motivation for these parameters and limitations of the adaptive optics system below. The transmitted beam had $w_0 = 10.6$ cm clipped at diameter $b = 15a = 30$ cm. The corresponding transfer function for the aperture and phase modulator during iteration $n$ is given by

$$\Theta^{(n)}(\mathbf{r}) = [1 - H(r - b/2)]\exp\left[-\frac{r^2}{w_0^2} + iu^{(n)}(\mathbf{r})\right], \qquad (10)$$

where $H$ is defined as the Heaviside step function and $u^{(n)}(\mathbf{r})$ is the phase applied by the phase modulator. Simulations were conducted for 300 statistically independent turbulent channels. Each simulation completed 100 iterations of the SPGD phase correction algorithm.

Figures (3a) and (3b) show the intensity profiles during iterations 1 and 100 respectively where the correction algorithm was most effective, i.e. producing the highest maximum intensity on the target. In this instance, the maximum intensity during the first iteration is 1.7 due to minor scintillation effects, but there does not exist a localized region of high intensity. After 100 iterations of the SPGD algorithm, the metric increased from $J = 304$ to $J = 1,280$. The beam during iteration 100, displayed in Fig (3b), exhibits a single localized region of high intensity, 9.5 times the transmitted beam's maximum intensity with approximately 10% of the transmitted power contained in a radius of 1 cm about the peak intensity. Out of the 300 simulations there were only 9 instances where the correction algorithm did not increase the maximum intensity on the target. Figures (3c) and (3d) display the intensity profiles during iterations 1 and 100 respectively for one such instance. In this simulation, the SPGD algorithm increased the metric from $J = 77$ to $J = 880$, while the maximum intensity decreased from 2.6 to 2.1. Instead of a single, localized region of high intensity, there are multiple local intensity maxima. By chance, the incoherent sums for each of these maxima added constructively, and resulted in an increased metric without an increased intensity on the target. In practice, the combination of a moving target with a finite photodetector exposure time would effectively perform an average and may mitigate such an occurrence.

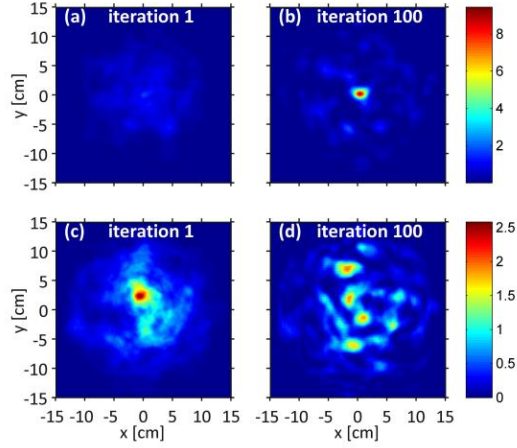

Figure 3. Top row: Initial (a) and final (b) intensity profiles in the turbulent channel where the SPGD correction algorithm was most effective. Bottom row: Initial (c) and final (d) intensity profiles in the turbulent channel where the SPGD correction algorithm was least effective.

Statistics over all 300 simulations are displayed in Fig. (4). Figures (4a) and (4b) display histograms of the maximum intensity and power in the bucket (PIB) on the target. The PIB is calculated as fractional power contained within a 1 cm radius bucket centered on the location of maximum intensity. Statistics obtained from intensity profiles during iterations 1 and 100 are plotted as the red and blue bars respectively. Due to minor scintillation, the average maximum intensity during iteration 1 is 1.9 with a standard deviation of .36. After 100 iterations of SPGD the average maximum intensity is 3.5 with a standard deviation of 1.2. The same trend is seen in the PIB. The average PIB during iteration 1 is .027 with a standard deviation of .005 and the average PIB during iteration 100 is .045 with a standard deviation of .015.

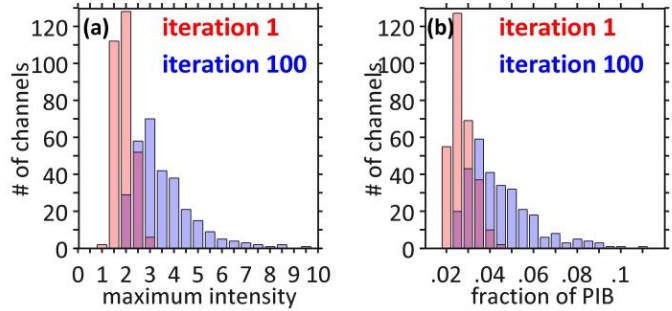

Figure 4. Histogram of maximum intensities (a) and fraction of received power within a bucket of radius 1cm (b) over 300 independent turbulent channels. The statistics corresponding to iteration 1 are denoted by red and the statistics corresponding to iteration 100 are denoted by blue.

In choosing the parameters for adaptive focusing based on Eq. (7), several conditions and limitations must be considered. First, there cannot be substantial intensity scintillation. Since our adaptive optics implementation only applies phase corrections, it cannot effectively correct for strong scintillation [20-22]. For weak turbulence, the scintillation index equals the Rytov variance, $\sigma_R^2 = 1.23 C_n^2 k^{7/6} z^{11/6}$, providing the condition $\sigma_R^2 < .3$. Recent research in combined amplitude and phase corrections may relax this condition [22]. Second, the dimension of the phase modulator elements must be less than the transverse coherence length, $a < \rho_0$, where $\rho_0 = 1.67(k^2 C_n^2 z)^{-3/5}$. Ideally a fast algorithm for correcting the wavefront would allow arbitrarily small phase modulator elements. However, with iterative algorithms, decreasing the element dimension can require decreasing the perturbation magnitude, $\varepsilon$, which increases the convergence time. Finally, iterative algorithms find local optima, which are not necessarily the ideal global optimum [19]. More sophisticated algorithms may offer a solution, but are a fundamental adaptive optics problem beyond the scope of this manuscript. Working within these limits, we have demonstrated the derived relation can facilitate the formation of a localized region of high intensity on a remote, spatially incoherent target. The derived relation is quite general, relying entirely on reciprocity, making it suitable for applications requiring propagation through random media.

The authors would like to thank J. Peñano, P. Sprangle, L. Andrews, and R. Phillips for fruitful discussions. This work was supported by funding from JTO through ONR under Contract No. N000141211029.